# Angular momentum projection of cranked Hartree-Fock states: Application to terminating bands in $A\sim44$ nuclei

H. Zduńczuk,[1] W. Satuła,[1] J. Dobaczewski,[1,2] and M. Kosmulski[3]

[1]*Institute of Theoretical Physics, University of Warsaw, ul. Hoża 69, 00-681 Warsaw, Poland.*
[2]*Department of Physics, P.O. Box 35 (YFL), FI-40014 University of Jyväskylä, Finland*
[3]*Department of Physics, University of Warsaw, ul. Hoża 69, 00-681, Warsaw, Poland*
(Dated: October 27, 2018)

We present the first systematic calculations based on the angular-momentum projection of cranked Slater determinants. We propose the $I_y\rightarrow I$ scheme, by which one projects the angular momentum $I$ from the 1D cranked state constrained to the average spin projection of $\langle \hat{I}_y\rangle = I$. Calculations performed for the rotational band in $^{46}$Ti show that the AMP $I_y\rightarrow I$ scheme offers a natural mechanism for correcting the cranking moment of inertia at low-spins and shifting the terminating state up by $\sim 2$ MeV, in accordance with data. We also apply this scheme to high-spin states near the band termination in $A\sim44$ nuclei, and compare results thereof with experimental data, shell-model calculations, and results of the approximate analytical symmetry-restoration method proposed previously.

## I. INTRODUCTION

The energy density functional (EDF) method in nuclear physics is nowadays the approach of choice for large-scale nuclear structure calculation. It has the same roots as the density functional theory (DFT) [1] in atomic and molecular physics, which is based on the Coulomb interaction between electrons. In nuclear physics, the EDF approach still lacks firm microscopic derivation and well-defined rules that would allow for systematic construction of an exact (or optimal) functional. In practice, the nuclear EDF is constructed phenomenologically, based on the knowledge accumulated within modern self-consistent mean-field approaches built upon effective density-dependent two-body interaction.

These approaches, although successful in reproducing gross nuclear properties and certain generic features of collective and single-particle nuclear motion are only coarse in confrontation with precise spectroscopic data. This situation calls for improvement, which can be achieved in two directions, namely, either *explicitly*, by using better and/or more complicated parameterizations of the nuclear EDF, or *implicitly*, by going beyond the mean-field or Hartee-Fock (HF) level. At present, intense studies are being conducted in both directions, aiming at exploring the limits thereof and answering the question of whether they can be considered equivalent, complementary, or independent. In this work we will explore the second alternative, by employing the angular-momentum-projection (AMP) method of cranked HF (CHF) states.

The spontaneous symmetry breaking (SSB) mechanism is inherently built into the HF approach. In many cases, it allows not only for incorporating a significant fraction of many-body correlations into a single HF state, but also serves as a source of deep physical intuition. Emergence of nuclear deformation (breaking of rotational invariance) leads naturally to the collective rotational motion, and is one of the most spectacular manifestations of the SSB in nuclear physics.

The CHF approximation treats collective (rotational) and intrinsic degrees of freedom on the same self-consistent footing. This fact is at the base of the success of this simple approximation, because in atomic nuclei both energy scales are strongly interwoven, and dramatic structural changes may take place along rotational bands. Terminating rotational bands [2] are the best examples of such possible changes, whereupon the collective rotation is followed by the total alignment of valence particles.

Although the energies of rotational states are correctly reproduced by the CHF approach, and the changes of shape and pairing as well as the recoupling processes of individual nucleonic orbits are well captured, the price paid is high. Indeed, the resulting wave packets (deformed CHF states) $|\Phi_{I_y}\rangle$ are well localized in the angular degrees of freedom and thus they are broadly spread over many angular momenta $|\Phi_{I_y}\rangle = \sum_I w_I |I\rangle$, with only the average value of the projection of angular momentum on one the axes (the $y$ axis in our case) being constrained,

$$\langle\Phi_{I_y}|\hat{I}_y|\Phi_{I_y}\rangle \equiv \langle\hat{I}_y\rangle = I_y. \tag{1}$$

Apart from strongly deformed states, this feature of the cranking approximation precludes applications of this formalism to compute transition rates, which constitute extremely valuable source of structural information. The demand for symmetry-restoration is therefore well motivated. Starting from deformed CHF state $|\Phi_{I_y}\rangle$, the goal can be achieved by projecting onto the eigenspaces of the angular momentum. Most of the calculations that have been performed so far invoke the angular momentum method applied to non-rotating states, see, e.g., Refs. [3, 4, 5] and the reviews in Refs. [6, 7, 8, 9]. This limits their applicability to low spin states, where the influence of rotation on the intrinsic states can be neglected, and leads to an overestimate of the nuclear moment of inertia (MoI) as compared to the (realistic) cranking estimate.

After the ground-breaking studies in Refs. [10, 11], the AMP of CHF states has not been performed in modern self-consistent calculations in nuclear structure. Recently, in Ref. [12], we presented results of such calculations for the test case of collective rotation of a well-deformed nucleus $^{156}$Gd. The AMP procedure we used has been implemented within the code HFODD [13, 14]. The calculational scheme proposed and tested in Ref. [12], dubbed hereafter $I_y{\rightarrow}I$ scheme, assumes the AMP of spin component $I$ of the self-consistent CHF state $|\Phi_{I_y}\rangle$, which is constrained to the same mean value of the projection on the $y$ axis, i.e., $\langle\hat{I}_y\rangle = I$. It combines the simplicity of the self-consistent 1D cranking approach, and its ability to reproduce the correct MoI, with the AMP after variation method.

In this paper, within the same formalism, we present the first systematic calculations of rotational states along terminating bands in the $A{\sim}44$ mass region. The paper is organized as follows. Methods of calculation and results are presented in Secs. II and III, respectively. In particular, the AMP of cranked states along the rotational band in $^{46}$Ti is discussed in Sec. III A and the AMP of states near the band termination is presented in Sec. III B. Finally, summary and discussion are given in Sec. IV.

## II. METHODS

The angular-momentum-conserving wave function is obtained by employing the standard operator $\hat{P}^I_{MK}$ [15, 16] projecting onto angular momentum $I$, with projections $M$ and $K$ along the laboratory and intrinsic $z$ axes, respectively,

$$|IMK\rangle = \hat{P}^I_{MK}|\Phi_{I_y}\rangle \equiv \frac{2I+1}{8\pi^2} \int D^{I*}_{MK}(\Omega)\,\hat{R}(\Omega)|\Phi_{I_y}\rangle\,d\Omega. \quad (2)$$

Here, $\Omega$ represents the set of three Euler angles $\alpha\beta\gamma$, $D^{I*}_{MK}(\Omega)$ are the Wigner functions [17], and $\hat{R}(\Omega) = e^{-i\alpha\hat{I}_z}e^{-i\beta\hat{I}_y}e^{-i\gamma\hat{I}_z}$ is the rotation operator.

Since $K$ is not a good quantum number, different $K$ components must be mixed with the mixing coefficients determined by the minimization of energy. The $K$-mixing is realized in a standard way by assuming:

$$|IM\rangle^{(i)} = \sum_K g^{(i)}_K |IMK\rangle \equiv \sum_K g^{(i)}_K \hat{P}^I_{MK}|\Phi\rangle, \quad (3)$$

and by solving the following Hill-Wheeler (HW) [18] equation:

$$\mathcal{H}\bar{g}^{(i)} = E_i \mathcal{N}\bar{g}^{(i)}, \quad (4)$$

where $\mathcal{H}_{K'K} = \langle\Phi|\hat{H}\hat{P}^I_{K'K}|\Phi\rangle$ and $\mathcal{N}_{K'K} = \langle\Phi|\hat{P}^I_{K'K}|\Phi\rangle$ denote the Hamiltonian and overlap kernels, respectively, and $\bar{g}^{(i)}$ denotes a column of the $g^{(i)}_K$ coefficients. The overlap and Hamiltonian kernels have their standard functional form but depend upon transition (or mixed) density matrices between rotated states:

$$\rho_{\alpha\beta}(\Omega) = \frac{\langle\Phi|a^+_\beta a_\alpha \hat{R}(\Omega)|\Phi\rangle}{\langle\Phi|\hat{R}(\Omega)|\Phi\rangle}. \quad (5)$$

The transition density matrix is also used for the density-dependent term. This is the only prescription available so far satisfying certain consistency criteria, formulated and thoroughly discussed in Refs. [5, 19].

Due to the overcompleteness of the $|IMK\rangle$ states, Eq. (4) is solved within the so-called collective subspace spanned by the natural states:

$$|mIM\rangle = \frac{1}{\sqrt{n_m}} \sum_K \eta^{(m)}_K |IMK\rangle, \quad (6)$$

which are eigenstates of the norm matrix $\mathcal{N}_{K'K}$ having non-zero eigenvalues ($n_m \neq 0$):

$$\mathcal{N}\bar{\eta}^{(m)} = n_m \bar{\eta}^{(m)}. \quad (7)$$

In practical applications, the cutoff parameter $\zeta$ is used and the collective subspace is constructed by using only those natural states that satisfy $n_m \geq \zeta$. By ordering indices $m$ of the natural states in such a way that larger indices correspond to smaller norm eigenvalues, we can write the solutions of the HW equation (4) as:

$$|IM\rangle^{(i)} = \sum_{m=1}^{m_{\max}} f^{(i)}_m |mIM\rangle, \quad (8)$$

where the mixing coefficients of Eq. (3) read:

$$g^{(i)}_K = \sum_{m=1}^{m_{\max}} \frac{f^{(i)}_m \eta^{(m)}_K}{\sqrt{n_m}}. \quad (9)$$

We can now define two types of the $K$-mixing. On the one hand, by the kinematic $K$-mixing we understand the situation where only one collective state is used, i.e., $m_{\max} = 1$. Then, the solution of the HW equation amounts to calculating only one matrix element of the Hamiltonian kernel,

$$E_1 = \bar{g}^{(1)\dagger}\mathcal{H}\bar{g}^{(1)} = \frac{\bar{\eta}^{(1)\dagger}\mathcal{H}\bar{\eta}^{(1)}}{n_1}, \quad (10)$$

i.e., $f^{(1)}_1 = 1$ and $f^{(1)}_m = 0$ for $m > 1$. In the kinematic $K$-mixing, the mixing coefficients $g^{(1)}_K = \eta^{(1)}_K/\sqrt{n_1}$ are entirely determined by the norm kernel and do not depend on the Hamiltonian kernel, i.e., they are entirely given by the cranking approximation and Coriolis coupling. On the other hand, by the dynamic $K$-mixing we understand the full solution of the HW equation for $m_{\max} > 1$, where the cutoff parameter $\zeta$ is adjusted so as to obtain a plateau condition for the lowest eigenvalue $E_1$. Here, the generator-coordinate-method (GCM) mixing of different $K$ components becomes effective, which potentially can modify the cranking mixing coefficients. We stress here that the kinematic $K$-mixing does correspond to a $K$-mixed solution too, and is not assuming any single given value of $K$.

The deformed CHF states were provided by the code HFODD, which solves the Hartree-Fock equations that correspond to the Ritz variational principle,

$$\delta\frac{\langle\Phi_{I_y}|\hat{H}-\omega\hat{I}_y|\Phi_{I_y}\rangle}{\langle\Phi_{I_y}|\Phi_{I_y}\rangle}=0, \quad (11)$$

with angular frequency $\omega$ adjusted so as to fulfill constraint (1) and the value of $I_y$ being equal to $I$, according to our $I_y\rightarrow I$ scheme. The $y$-signature and parity symmetries were conserved. The Hamiltonian and overlap kernels were calculated using the Gauss-Chebyshev quadrature in the $\alpha$ and $\gamma$ directions and the Gauss-Legendre quadrature in the $\beta$ direction [20]. In the numerical applications presented in this work we used the SLy4 [21] Skyrme force, but similar results were also obtained by using the SIII [22] force. The time-odd terms in the Skyrme functional were fixed by using values of the Landau parameters [23, 24]. The harmonic-oscillator basis was composed of 10 spherical shells. The integration over the Euler angles was done by using a cube of $50\times50\times50$ integration points.

## III. RESULTS

In Ref. [12], we demonstrated that in a well-deformed nucleus $^{156}$Gd, the AMP $I_y\rightarrow I$ scheme reproduces well the cranking MoI along the rotational band, after taking into account the dynamic $K$-mixing. Here, we study the AMP methods applied to the terminating bands in the $A\sim44$ region of nuclei.

### A. Ground-state rotational band in $^{46}$Ti

As a first example, consider the ground-state rotational band in $^{46}$Ti. According to the CHF model, the shape of this nucleus undergoes a gradual change along the band. Starting from a well elongated ($\beta_2\sim0.23$) shape at low spins, the nucleus goes through the alignment processes of the $f_{7/2}$ protons and neutrons, and eventually reaches a nearly spherical ($\beta_2\sim0.05$) shape at the terminating spin of $I=14$, as shown in Fig. 1a.

The calculated and experimental $^{46}$Ti rotational bands are shown in Fig. 1b. Results of the CHF calculations (left) are compared to those of AMP (middle) and to experimental data (right). In order to visualize the role of the $K$-mixing, we have depicted results of the AMP calculations separately for the kinematic and dynamic $K$-mixing. For the dynamic $K$-mixing, the AMP excitation energies were calculated from the plateau condition. The stability of results with respect to the number of natural states $m_{\max}$ is shown in Fig. 1c. It can be seen that already at $m_{\max}=2$, a perfect stability is obtained, i.e., here, the difference between the kinematic and dynamic $K$-mixing is related to adding the $m=2$ state to the collective subspace. At higher values of $m_{\max}$, one observes small departures from the converged values, which are due to accidental mixing with spurious solutions (the largest such an effect is seen for $I=4$ at $m_{\max}=5$). Nevertheless, physical converged solutions are clearly seen well beyond the point where the spurious solutions become lower in energy, as is well known for other generator-coordinate-method calculations [26].

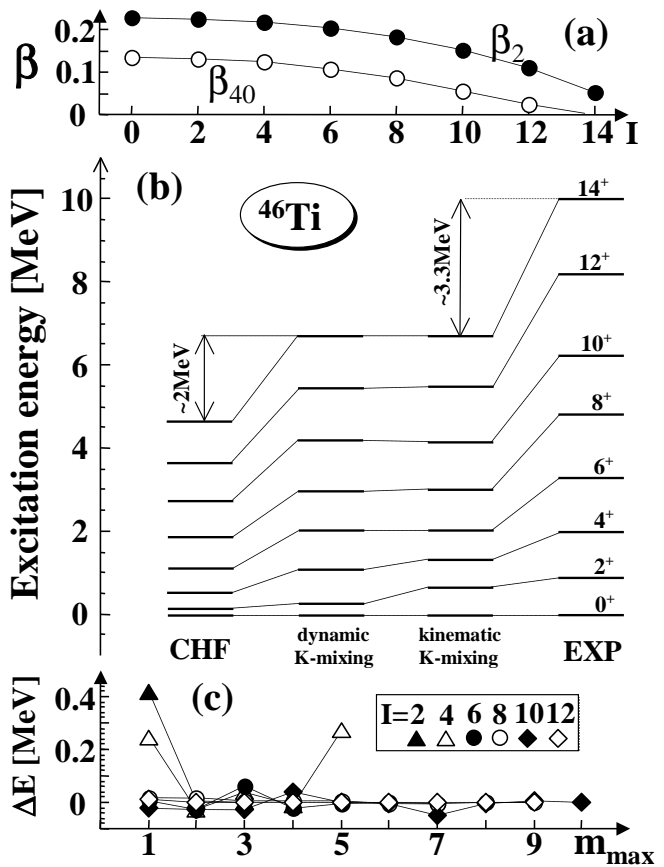

FIG. 1: Results of the CHF and AMP calculations for the rotational band in $^{46}$Ti obtained for the SLy4 interaction. The upper panel (a) shows the evolution of quadrupole and hexadecapole shape parameters along the band as calculated using CHF approximation. The middle panel (b) shows the excitation spectra calculated using the CHF approximation (left column) and AMP method with kinematic and dynamic $K$-mixing (middle columns), compared with the empirical data [25] (right column). The lower panel (c) shows deviations $\Delta E$ of energy levels with spins $I=2,\ldots,12$ from the converged values corresponding the plateau condition, as functions of the number $m_{\max}$ of the norm eigenvalues used when solving the HW equation.

Figure 1b clearly shows that within the $I_y\rightarrow I$ scheme, the AMP effectively causes a decrease of the mean MoI within the band. This effect is expected to be generic for rotational bands of decreasing collectivity. Indeed, in such cases rotational correction is large at low spins and decreases at higher spins. This is illustrated in Fig. 2, where the calculated bands are shown in the absolute energy scale. In the case of $^{46}$Ti, a net increase in the excitation energy of the terminating state, due to the AMP, amounts to about 2 MeV, but it is almost entirely related to the lowering of the $I = 0$ state, while the terminating $I = 14$ state remains almost unaffected by the AMP.

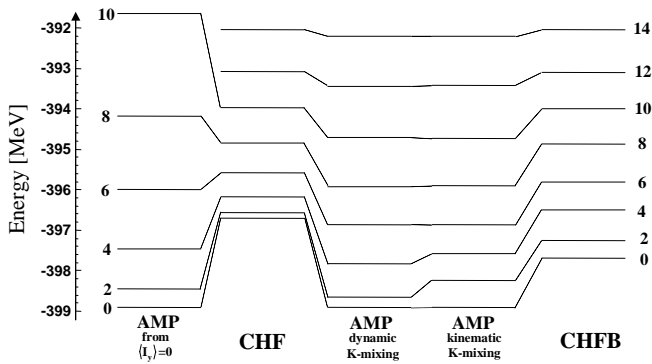

FIG. 2: Similar as in Fig. 1b, but shown in the absolute energy scale. Results of the AMP from the $\langle \hat{I}_y \rangle = 0$ ground state (first column) are compared to those of the CHF (second column), AMP with kinematic and dynamic $K$-mixing (third and fourth columns), and CHFB calculations (fifth column).

In the same Figure, we also show for comparison the AMP spectrum obtained by projecting from the $\langle \hat{I}_y \rangle = 0$ ground state. In this case, the overall MoI turns out to be much too small and the excitation energy much too high as compared to the CHF or AMP $I_y \to I$ results. This fact once again shows the importance of the structural changes that occur in the system with increasing angular momentum.

Note that the $\langle \hat{I}_y \rangle = 0$ state is axial, and thus for all spins contains only the $K = 0$ component. Therefore, here, the collective space contains only one state and there is no $K$-mixing at all. Moreover, axial symmetry leads to a tremendous simplification of the AMP method, whereby only one-dimensional integration over the Euler $\beta$ angle is needed. Note also that the CHF states for $\langle \hat{I}_y \rangle \neq 0$ are never axial, because the Coriolis coupling always induces some non-zero nonaxiality. Therefore, the full three-dimensional integration over the Euler angles $\alpha\beta\gamma$ is needed in our $I_y \to I$ scheme.

In the right column of Fig. 2, we also show results of the cranked Hartree-Fock-Bogolyubov calculations (CHFB) performed by using a zero-range volume-type (density-independent) interaction in the pairing channel. Its strengths of $V_n = -217.0$ and $V_p = -237.5$ MeV fm$^{-3}$ for neutrons and protons, respectively, with the cutoff energy of $\epsilon_{\text{cut}} = 50$ MeV [27], was adjusted so as to reproduce, on average, the pairing gaps in this region of nuclei. For the $\langle \hat{I}_y \rangle = 0$ ground state of $^{46}$Ti, the calculated gaps read $\Delta_n = 1.394$ and $\Delta_p = 1.632$ MeV.

One can see that, in comparison with the CHF results, pairing correlations only affect states at low spins, $I \leq 6$, and give about 1 MeV of additional binding at $\langle \hat{I}_y \rangle = 0$. Pairing correlations vanish in the terminating state, and thus its energy is the same within the CHF and CHFB methods. At present, the code HFODD cannot perform the AMP of paired states, and we are yet unable to evaluate combined effects of pairing and AMP from cranked states.

It is also interesting to observe that, in the case of $^{46}$Ti, the dynamic $K$-mixing is effective essentially only for $I = 2$ and 4, see Fig. 1b,c. This result is rather surprising, particularly in view of the fact that the only discontinuity in spatial anisotropy of the CHF solutions can be seen around spins $I \sim 6, 8$. This behavior is also qualitatively different from that found in $^{156}$Gd [12], where the dynamic $K$-mixing was effective up to the highest ($I \sim 20$) calculated spins. Apparently, the magnitude of $K$-mixing cannot be inferred solely from the shape but it also depends on individual (alignment) degrees of freedom.

In Fig. 3, we show probabilities $W_I$ of finding the angular-momentum components in the intrinsic CHF

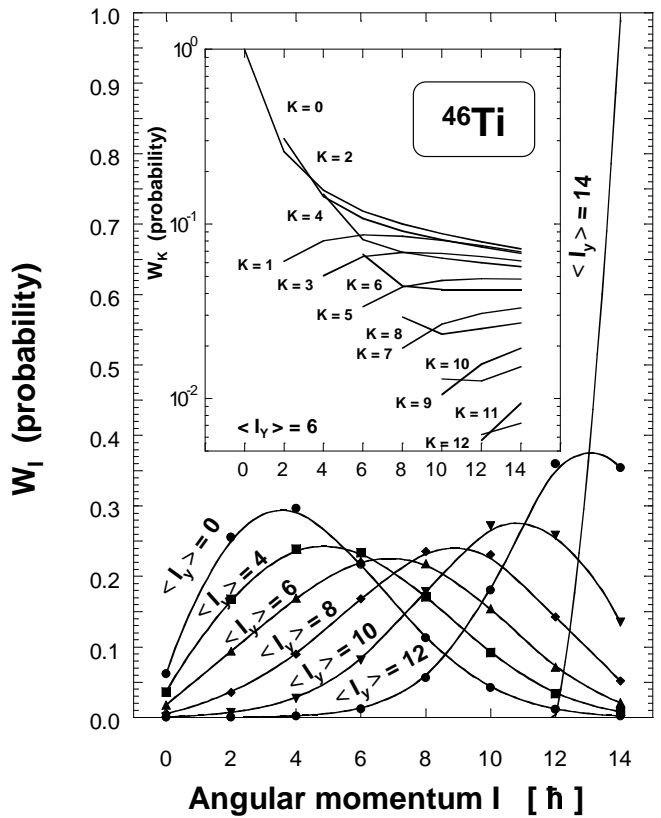

FIG. 3: Probabilities $W_I$ (12) and $W_K$ (13) of finding the given $I$ and $K$ components, respectively, in the intrinsic CHF states for different values of the projection on the $y$ axis $\langle \hat{I}_y \rangle$. Solid lines are drawn only to guide the eye. Note that due to the conserved $y$-signature symmetry, one has $W_I = 0$ for odd values of $I$ and $W_{-K} = W_K$.

states, i.e.,

$$W_I = \sum_{K=-I}^{I} \langle \Phi_{I_y} | \hat{P}_{KK}^I | \Phi_{I_y} \rangle. \quad (12)$$

The inset also shows probabilities $W_K$ of finding the given $K$ components,

$$W_K = \langle \Phi_{I_y} | \hat{P}_{KK}^I | \Phi_{I_y} \rangle / W_I, \quad (13)$$

in one of the intrinsic states, $\langle \hat{I}_y \rangle = 6$. At low spins, the $W_I$ probability distributions follow the general pattern of deformed collective states. This pattern suddenly disappears at the termination point of $\langle \hat{I}_y \rangle = 14$, where the CHF state becomes totally aligned, and therefore, it contains only the single $I = 14$ component. The $W_K$ probabilities do not follow the standard pattern of collective rotation, where they correspond to the Coriolis mixing only, cf. Fig. 2 in Ref. [12]. Here, the aligning state causes the values of $W_K$ to decrease with $I$ and induces the crossing of those corresponding to even and odd $K$ values in function of $I$. Note also, that the angular momentum aligns along the $y$ axis, while the standard projections $K$ are calculated with respect to the $z$ axis. Therefore, we do not expect to see a single $K$ component of the aligned $\langle \hat{I}_y \rangle = 14$ state.

We conclude this section by analyzing excitation spectra calculated using the AMP of CHF states constrained to different values of $\langle \hat{I}_y \rangle$, as shown in Fig. 4. It turns out that the AMP energies minimized with respect to $\langle \hat{I}_y \rangle$, or equivalently with respect to the angular frequency $\omega$, significantly differ from those obtained within our $I_y \to I$ scheme. Indeed, although the minimum $I = 0$ energy is obtained by projecting the $\langle \hat{I}_y \rangle = 0$ CHF state, this is not the case for $I > 0$ states. For example, the minimum $I = 2$ and 14 energies are obtained by projecting the $\langle \hat{I}_y \rangle = 0$ and 10 CHF states, respectively. The resulting excitation spectrum (in Fig. 4 indicated by arrows) is quite irregular and can hardly be associated with the physical result. The reason is the fact that the minimization over $\omega$ does not constitute the full variation-after-projection minimization of energy, which should be performed in the complete parameter space, including, e.g., the deformation parameters. Moreover, the fact that the $I = 14$ state may gain energy by using collective correlations on top of the fully aligned (terminating) state points out a possible inadequacy of the EDF method when it is used within the variation-after-projection procedure, and not within the $I_y \to I$ projection-after-variation scheme.

### B. Angular-momentum projection near the band termination

In the vicinity of band termination, the number of contributing configurations drops down and the physics simplifies significantly. Reliable approximate analytical symmetry restoration schemes can be easily derived for these cases; for details we refer the reader to the analysis presented recently in Refs. [28, 29]. In the present paper we aim at further studying and testing these approximate methods against rigorous AMP results.

Let us first consider the energy splittings between the favored- and unfavored-signature terminating states, $[f_{7/2}^n]_{I_{max}}$ and $[f_{7/2}^n]_{I_{max}-1}$, respectively, within the $[f_{7/2}^n]$ configurations, where $n$ denotes the number of particles in the $f_{7/2}$ sub-shell. Within the *naïve* noncollective cranking model, the unique aligned $|\Phi_{I_{max}}\rangle$ states can be considered as many-body reference states (HF vacua) with projections of the angular momentum being conserved quantum numbers equal to the maximum allowed values, $I_y = I_{max}$. From these local HF vacua, the $|\Phi_{I_{max}-1}^{(\tau)}\rangle$ states can be generated by particle-hole (ph) excitations; in particular, by changing either the signature of a single neutron ($\tau = \nu$) or a single proton ($\tau = \pi$). In spite of the fact that the underlying CHF solutions are almost spherical, they manifestly break the rotational invariance. Indeed, the two $I_{max}-1$ CHF solutions have conserved projections of the angular momentum, $I_y = I_{max}-1$, but are in this case linear combinations of the total-angular momentum states with $I = I_{max}$ and $I_{max}-1$, i.e., up to a normalization factor:

$$|\Phi_{I_{max}-1}^{(\pi)}\rangle \sim b|I_{max}; I_{max}-1\rangle + a|I_{max}-1; I_{max}-1\rangle,$$
$$|\Phi_{I_{max}-1}^{(\nu)}\rangle \sim a|I_{max}; I_{max}-1\rangle - b|I_{max}-1; I_{max}-1\rangle. \quad (14)$$

The simplicity of the encountered situation allows for an approximate analytical estimate of mixing coefficients $a$ and $b$ [28, 29]. Equivalently, one can find these coefficients by performing the exact AMP of the $|\Phi_{I_{max}-1}^{(\pi)}\rangle$ or $|\Phi^{(\pi)}_{I_{\max}-1}\rangle$ states. The resulting probabilities $P(I_{\max}-1)$ of finding the $I_{\max}-1$ components within the $|\Phi^{(\pi)}_{I_{\max}-1}\rangle$ states are shown in Fig. 5a. The AMP results match perfectly the analytical results obtained in Refs. [28, 29], confirming reliability of the approximate method.

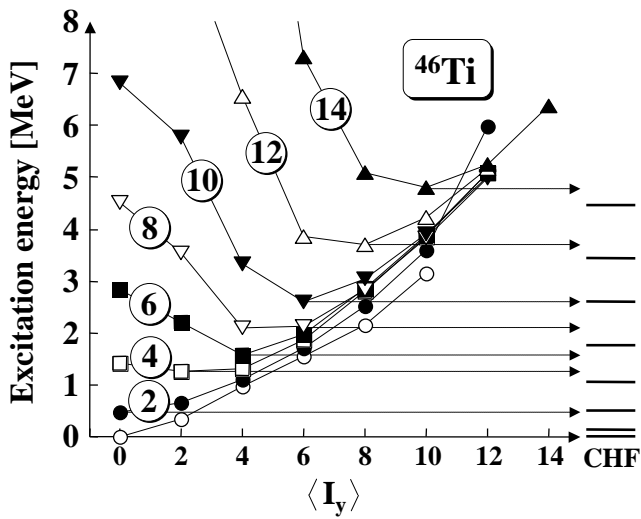

FIG. 4: Excitation spectra calculated using the AMP of CHF states constrained to different values of $\langle \hat{I}_y \rangle$, with kinematic $K$-mixing. Solid lines connect states of given projected angular momenta $I$ and are drawn only to guide the eye. Arrows indicate the minimum energies in function of $\langle \hat{I}_y \rangle$. For comparison, the right column shows the CHF spectrum.

Calculated energy differences $E(I_{\max}) - E(I_{\max}-1)$ are shown in Fig. 5b. Since the CHF solutions break the isobaric invariance, the AMPs of the $|\Phi^{(\pi)}_{I_{\max}-1}\rangle$ and $|\Phi^{(\nu)}_{I_{\max}-1}\rangle$ states are not fully equivalent, and lead to slightly different energies. Results shown in Fig. 5 represent arithmetic averages of both AMP energies. The only exception is $^{42}$Sc, where we were able to perform numerical integration with a desired accuracy only when projecting from the $|\Phi^{(\nu)}_{I_{\max}-1}\rangle$ CHF state, and the depicted point represents this single result. It is evident from the Figure that, except for $^{42}$Sc and $^{45}$Sc, the quality of the results is comparable to the state-of-the-art shell-model calculations of Refs. [29, 30].

The unfavored-signature $I_{\max}-1$ CHF states discussed above were created by building the signature-inverting ph excitations on top of the $I_{\max}$ reference state. Similar procedure applied to create $I_{\max}-2$ CHF solutions leads to several nearly-spherical states, located, on average, above the reference state.

In this case, however, the SSB mechanism enters the game by pushing the collective (deformed) CHF solution down, below the reference state, and relatively close to the empirical energy. Hence, by going from the $I_{\max}-1$ to $I_{\max}-2$ states, the physics changes quite dramatically, showing clearly two contrasting facets of the SSB mechanism of rotational symmetry in nuclear physics.

In the case of the $I_{\max}-1$ states, the symmetry restoration results in a repulsion of two nearly-degenerate proton and neutron CHF states, which are located above the reference $I_{\max}$ state. On the one hand, the reorientation mode, $|I_{\max}; I_{\max}-1\rangle$, is shifted down and becomes degenerate with the $|I_{\max}; I_{\max}\rangle$ solution, as required by the rotational invariance. On the other hand, the physical mode, $|I_{\max}-1; I_{\max}-1\rangle$, is shifted up and becomes the unfavored-signature terminating state.

In the case of the $I_{\max}-2$ states, the SSB mechanism results in a repulsion of several nearly-degenerate proton and neutron ph states. The collective CHF mode, $|\Phi_{I_{\max}-2}\rangle$, is shifted below the reference $I_{\max}$ state, in accordance with data. By the AMP mixing, the symmetry-restored collective mode projected from $|\Phi_{I_{\max}-2}\rangle$, i.e., the $|I_{\max}-2; I_{\max}-2\rangle$ state, gains some additional binding energy. The situation described above is schematically illustrated in the inset of Fig. 6.

Calculated energy differences, $E(I_{\max}) - E(I_{\max}-2)$, are shown in Fig. 6. The CHF solutions, except for $^{42}$Ca and $^{42,43}$Sc, correspond to collective states having $\beta_2 \sim 0.10 - 0.12$. The AMP shifts these states almost uniformly down by about 300 – 400 keV, enlarging the splitting by that amount, and improving an overall agreement between theory and experiment. It is, however, evident from the Figure that, here, the AMP does not improve upon the incorrect isotopic/isotonic dependence of the CHF results. The magnitude of rotational correction is determined predominantly by the shape change, and does not vary from case to case. One can speculate that a detailed agreement with data would require an additional isospin-symmetry restoration.

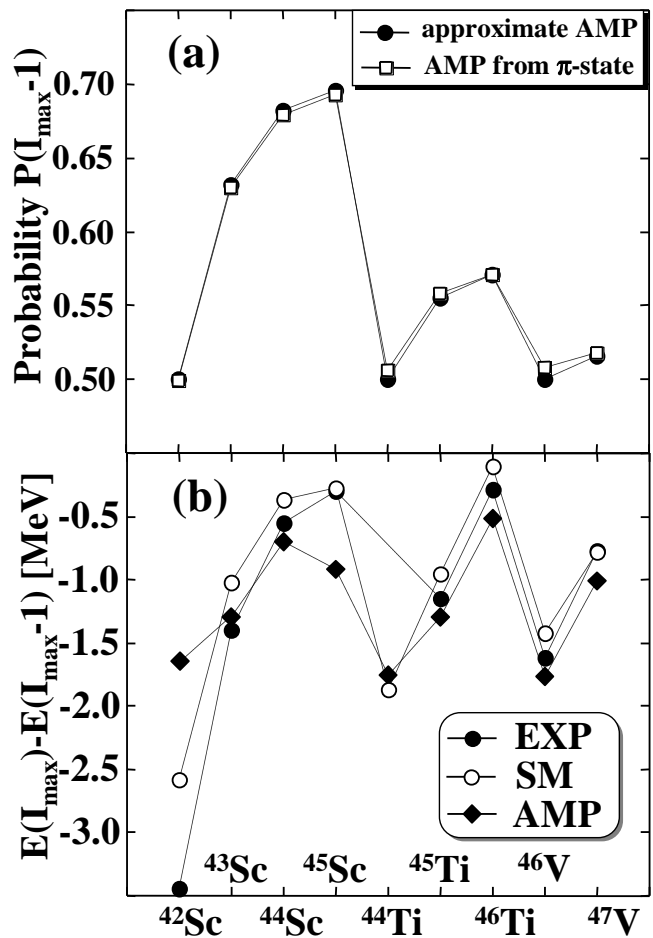

FIG. 5: Upper panel (a): probabilities of finding the $I_{\max}-1$ spin components in the $|\Phi^{(\pi)}_{I_{\max}-1}\rangle$ CHF solutions. Squares and dots show results calculated using the exact AMP and approximate method of Ref. [28], respectively. Lower panel (b): energy differences between the favored- and unfavored-signature terminating states, $[f^n_{7/2}]_{I_{\max}}$ and $[f^n_{7/2}]_{I_{\max}-1}$, respectively. Values calculated using the AMP method (diamonds) are compared to those obtained within the shell-model [28] (circles) and empirical data [31, 32, 33, 34, 35, 36] (dots).

A similar trend was found for the $I_{\max}-2$ states for configurations involving one-proton ph excitation from $d_{3/2}$ to $f_{7/2}$ subshell, see Fig. 7. Here, all the lowest $I_{\max}-2$ states are found to be deformed. The energy gain due to the AMP is again of the order of 300 – 400 keV and weakly depends on $N$ and $Z$, thus merely reflecting an increase of deformation between nearly-spherical $I_{\max}$ states and terminating collective cranking solutions for $I_{\max}-2$. This example confirms that the onset of col-lectivity causes a uniform mixing of various angular momenta, regardless of individual features of specific nuclei. Indeed, in the studied nuclei, probabilities of finding the $I_{\max}-2$ spin component within the $|\Phi_{I_{\max}-2}\rangle = \sum_I w_I |I\rangle$ CHF deformed wave packets are $|w_{I_{\max}-2}|^2 = 0.34 \pm 0.05$, i.e., they extremely weakly depend on $N$ and $Z$. Moreover, values of $|w_{I_{\max}-2}|^2$ appear to be very similar to $|w_{I_{\max}}|^2$, which are equal to $0.35 \pm 0.04$, and again almost do not change from one nucleus to another.

Situation changes quite radically for the unfavored-signature $[d_{3/2}^{-1} f_{7/2}^{n+1}]_{I_{\max}-1}$ states. Within the CHF approximation, in $N \neq Z$ and $N \neq Z+1$ nuclei there are three $I_{\max}-1$ configurations that can be created from the $I_{\max}$ reference state. Indeed, this can be done by a signature-inverting ph excitation involving either the neutron ($\nu$) or proton ($\pi$) $f_{7/2}$ particle, or the proton $d_{3/2}$ hole ($\bar\pi$), see Ref. [28]. The CHF solutions represent, therefore, mixtures of two physical $|I_{\max}-1; I_{\max}-1\rangle_i$, $i = 1, 2$, states with the spurious reorientation $|I_{\max}; I_{\max}-1\rangle$ mode.

In such a case, the $I_y \to I$ AMP scheme only removes the spurious mode, not affecting the mixing ratio of the two physical solutions $|I_{\max}-1; I_{\max}-1\rangle_i$, $i = 1, 2$. Hence, in contrast with the case of the $[f_{7/2}^n]_{I_{\max}-1}$ states, quality of the results strongly depends on the quality of the underlying CHF field. The AMP results corresponding to the $\bar\pi$ CHF solutions, which in $^{42-45}$Sc, $^{44-46}$Ti, and $^{47}$V are the lowest in energy, show that the admixtures of spurious components are of the order of 10%, see the inset in Fig. 8. The obtained rotational corrections are, therefore, small — of the order of $100-200$ keV, and the disagreement with data remains quite large, as shown in Fig. 8. We show these results only as an example of possible AMP calculations. However, for a complete analysis, one should, in principle, perform the GCM mixing of the AMP states corresponding to any possible CHF $I_{\max}-1$ configuration. A study in this direction is left for the future work.

## IV. SUMMARY AND DISCUSSION

The present work reports on the first systematic calculations using the AMP of cranked Hartree-Fock states. The technique used, called the $I_y \to I$ projection scheme, assumes the projection of the angular momentum component $I$ from the one-dimensional (1D) cranked Hartree-Fock solution constrained to $\langle \hat I_y \rangle = I$. The method benefits naturally from such nice physical features of the 1D cranking model as the shape-spin self-consistency or the ability to get a realistic estimate of the nuclear MoI. It is shown that the $I_y \to I$ AMP scheme leads to values of MoI that are much more realistic than those obtained by using the AMP of non-rotating $\langle \hat I_y \rangle = 0$ states, as was the common practice up to now.

In particular, application of the scheme to the rotational band in $^{46}$Ti clearly improves the MoI at the bottom of the band. It also reveals a simple mechanism by which rotational corrections allow for improving upon an incorrect excitation energy of the terminating state, $\Delta E_{I_{\max}}$, obtained in the CHF calculations, which account only for roughly half of the empirical value.

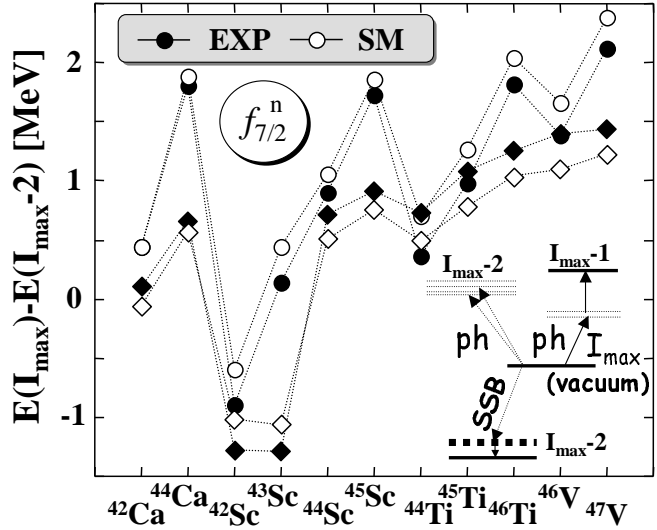

FIG. 6: Energy differences between the $[f_{7/2}^n]_{I_{\max}}$ terminating states and the lowest $[f_{7/2}^n]_{I_{\max}-2}$ states. Dots and circles represent the empirical data [31, 32, 33, 34, 35, 36, 37] and SM results of Ref. [28], respectively. Open diamonds label the lowest CHF solutions, which are collective except for the cases of $^{42}$Ca and $^{42,43}$Sc. Full diamonds represent the AMP results. Relative energies between the $I_{\max}$ reference state, non-collective ph $I_{\max}-1$ and $I_{\max}-2$ excitations (thin lines), SSB effect in the $I_{\max}-2$ state (thick dashed line), and final AMP configuration mixing (solid lines) are schematically illustrated in the inset.

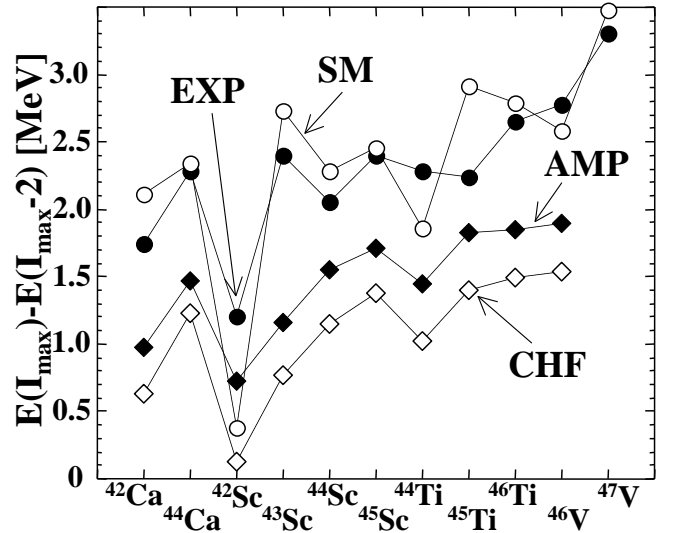

FIG. 7: Energy differences between the $[d_{3/2}^{-1} f_{7/2}^{n+1}]_{I_{\max}}$ terminating states and the lowest $[d_{3/2}^{-1} f_{7/2}^{n+1}]_{I_{\max}-2}$ states. Dots and circles represent empirical data [31, 32, 33, 34, 35, 36, 37] and SM results of Ref. [28], respectively. Open and full diamonds label the collective CHF solutions and AMP results, respectively.

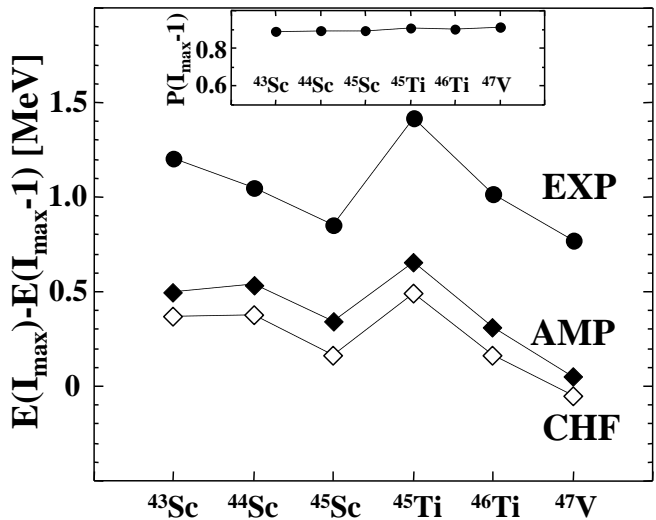

FIG. 8: Energy differences between the $[d_{3/2}^{-1} f_{7/2}^{n+1}]_{I_{\max}}$ terminating states and the lowest unfavored-signature terminating $d_{3/2}^{-1} f_{7/2}^{n+1}]_{I_{\max}-1}$ states. Dots represent empirical data [32, 33, 34, 35, 36, 37] and open and full diamonds represent the CHF and AMP results, respectively. Inset shows probabilities of finding the $I_{\max}-1$ components within the lowest CHF solution $|\Phi_{I_{\max}-1}\rangle$.

Pairing correlations that are active in the ground state, but do not affect fully aligned terminating state, can heal the situation only partially. Indeed, there is an upper limit for pairing correlation energy in the ground state, which can be sustained by a deformed $A{\sim}44$ system, equal to about 2 MeV. Further enhancement of pairing would induce the phase transition from deformed to spherical shape [38], and the rotational band could not have been built, contradicting the experimental data. Together, the rotational and pairing effects can bring $\Delta E_{I_{\max}}$ to about 9 MeV, i.e., some $\sim 10\%$ below the experimental value. Let us mention here that for the $I = 14$ state projected from the $\langle \hat{I}_y \rangle = 0$ ground state one obtains $\Delta E_{I_{\max}} \simeq 19$ MeV, i.e., the result, which is well above the empirical excitation energy of the terminating state.

For nearly-spherical unfavored-signature $[f_{7/2}^n]_{I_{\max}-1}$ terminating states, our AMP calculations give results in excellent agreement with data, and validate approximate projection methods introduced in Refs. [28, 29]. We also show that the onset of collectivity in the $[f_{7/2}^n]_{I_{\max}-2}$ states is quite correctly reproduced by the CHF calculations, on top of which the AMP gives only a small correction going in the right direction in comparison with data. However, details of the isotopic dependence are not reproduced here.

Similar conclusions are obtained for the $[d_{3/2}^{-1} f_{7/2}^{n+1}]$ configurations that involve one-proton ph excitation across the $Z = 20$ shell gap. In this case, in both $I_{\max}-1$ and $I_{\max}-2$ states near the band termination the collectivity sets in, while the energy differences with respect to terminating $I_{\max}$ states are underestimated in the CHF and AMP calculations.

The AMP of cranked Hartree-Fock states presented in this work was performed by applying the standard projection techniques to the CHF solutions obtained within the EDF method. We have checked that all the results are stable with respect to numerical parameters such as, e.g., numbers of integration points used when integrating kernels over the Euler angles. Nonetheless, one should be aware of potential risks caused by difficult to control, uncompensated poles plaguing projection techniques of states obtained within the EDF method [12, 39]. Clearly, the future of projection methods crucially depends on a satisfactory solution of the problem of such singularities.

This work was supported in part by the Polish Ministry of Science and by the Academy of Finland and University of Jyväskylä within the FIDIPRO programme.